\documentclass[twocolumn,twocolappendix]{aastex631}
\usepackage{bm}
\usepackage[T1]{fontenc}

\newcommand{\reviseone}{} 
\newcommand{\revisetwo}{} 
\newcommand{\revisethree}{} 

\received{October 13, 2025}
\accepted{November 12, 2025}

\submitjournal{ApJ}

\shorttitle{Contact Surfaces Formed by Pebble Collisions} 
\shortauthors{Tatsuuma et al.}

\begin{document} 

\title{Modeling the Contact Surfaces Formed by Pebble Collisions: Application to Formation of Comet 67P/Churyumov--Gerasimenko}


\correspondingauthor{Misako Tatsuuma}
\email{misako.tatsuuma@gmail.com}

\author[0000-0003-1844-5107]{Misako Tatsuuma}
\affiliation{Department of Electrical Engineering, National Institute of Technology, Tokyo College, 1220-2 Kunugida-machi, Hachioji, Tokyo 193-0997, Japan}
\affiliation{Division of Fundamental Mathematical Science, RIKEN Center for Interdisciplinary Theoretical and Mathematical Sciences (iTHEMS), 2-1 Hirosawa, Wako, Saitama 351-0198, Japan}

\author[0000-0002-1886-0880]{Satoshi Okuzumi}
\affiliation{Department of Earth and Planetary Sciences, Institute of Science Tokyo, 2-12-1 Ookayama, Meguro, Tokyo 152-8551, Japan}

\author[0000-0003-4562-4119]{Akimasa Kataoka}
\affiliation{Division of Science, National Astronomical Observatory of Japan, 2-21-1 Osawa, Mitaka, Tokyo 181-8588, Japan}

\author[0000-0001-9659-658X]{Hidekazu Tanaka}
\affiliation{Astronomical Institute, Graduate School of Science, Tohoku University, 6-3 Aramaki, Aoba-ku, Sendai 980-8578, Japan}

\begin{abstract} 

Modeling the contact surfaces formed by pebble collisions is crucial to understanding the formation process of comets, which are thought to be composed of pebbles.
In this paper, we develop a new model to estimate the contact surface radius and the number of contact points as functions of collision velocity, and examine the formation process of comet 67P/Churyumov--Gerasimenko.
Our model is based on the compressive strength of dust aggregates obtained from numerical simulations and assumes that all the impact energy of the pebbles is used for their mutual compression.
We compare our model with numerical simulations of pebble collisions, in which we prepare the initial pebbles in the form of compressed dust aggregate spheres and measure the contact surface and pebble radii using two- and three-dimensional characteristic radii, respectively.
We also apply our model to the formation scenario of comet 67P, whose tensile strength and bulk density have already been estimated in the literature.
We find that its low tensile strength points to formation via pebble collisions at velocities below $\sim10\mathrm{\ cm\ s^{-1}}$ when a microscopic filling factor of pebbles is lower than 0.6, suggesting that inelastic bouncing collisions played a role in damping the collision velocities.
By assuming that the pebble collision velocity is determined by the transition velocity between bouncing and sticking, we estimate the pebble radius inside comet 67P to be 130 $\mathrm{\mu m}$ or smaller.

\end{abstract}

\keywords{Comets (280), Planetesimals (1259), Planet formation (1241), Computational methods (1965), Analytical mathematics (38)}

\section{Introduction} \label{sec:intro}

The formation process of kilometer-sized planetesimals begins with the coagulation of submicrometer-sized dust grains and the compression of dust aggregates.
The growth of aggregates is halted by bouncing and fragmentation, preventing direct planetesimal formation through sticking \citep[e.g.,][]{Weidling2009,Guttler2010,Zsom2010,Krijt2018,Schrapler2022}.
Observations of protoplanetary disks suggest that submillimeter- to centimeter-sized dust aggregates are commonly present and are known as ``pebbles'' \citep[e.g.,][]{Perez2012,Perez2015,Testi2014,Kataoka2016HLTau,Kataoka2016HD,Kataoka2017,Tazzari2016,Bacciotti2018,Carrasco-Gonzalez2019,Liu2019,Miotello2023,Drazkowska2023}.
These pebbles concentrate due to the streaming instability and subsequently undergo gravitational collapse to form planetesimals \citep[e.g.,][]{Goldreich1973,Youdin2005,Johansen2007,Johansen2011}.
{\reviseone{The planetesimal formation process from pebbles depends on the physical properties of pebbles.}}

{\reviseone{Comets, which are considered largely unprocessed remnants of planetesimals, offer unique insights into the building blocks and formation mechanisms of ice planetesimals.}}
Observations of comets suggest that they consist of pebbles.
For instance, the low tensile strength of comet 67P/Churyumov--Gerasimenko \citep[hereafter 67P; e.g.,][]{Groussin2015,Basilevsky2016,Hirabayashi2016,Attree2018} indicates that it is not a homogeneous body of dust aggregates but instead a loosely packed structure \citep[e.g.,][]{Skorov2012,Blum2014,Blum2017,Tatsuuma2019,Blum2022}.
Additionally, Rosetta's observations revealed millimeter- to centimeter-sized structures on its surface \citep[e.g.,][]{Poulet2016}, supporting the hypothesis that the comet may be composed of pebbles.
However, whether comets indeed formed via the gravitational collapse of pebble clouds is still under debate \citep[e.g.,][]{Johansen2014,Wahlberg2014,Lorek2016,Visser2021}.

To connect the gravitational collapse scenario of pebble clouds with cometary observations, it is essential to model the contact surfaces between pebbles.
The properties of the contact surfaces are determined by the collision velocity between pebbles during gravitational collapse.
The contact surfaces are inferred from observations of the tensile strength of comets \citep[e.g.,][]{Skorov2012,Blum2017}.
By relating the inferred contact surfaces to expected collision velocities, we can assess the plausibility of the pebble cloud collapse scenario for comet formation.

{\revisetwo{However, some previous estimates of the contact surfaces between pebbles have been based on macroscopic parameters of dust aggregates derived from experiments \citep{Gundlach2012,Weidling2012,Arakawa2023pebble}, and their dependence on the microscopic filling factor of pebbles and their composition remains unclear.
These contact surfaces are expected to depend on such microscopic parameters, and the implications for comet formation can be highly sensitive to these parameters.}}
Therefore, in this work, we propose a new model for the contact surfaces between pebbles, incorporating the compressive strength of dust aggregates derived from microscopic modeling \citep{Tatsuuma2023} and validate it through numerical simulations.
Our model offers an improved framework for interpreting cometary observations and constraining the planetesimal formation process.

The structure of this paper is as follows.
Section \ref{sec:setting} describes our simulation setup for numerically calculating the contact surfaces between colliding pebbles.
Section \ref{sec:model} presents our newly developed contact surface model.
Section \ref{sec:result} compares the simulation results with the model predictions.
Section \ref{sec:apply} discusses the implications for comet 67P formation.
Finally, Section \ref{sec:conclusion} summarizes our conclusions.

\section{Simulation Settings} \label{sec:setting}

In this section, we describe our simulation model for calculating the contact surfaces between colliding pebbles, consisting of aggregates of submicrometer-sized grains called monomers.
Section \ref{subsec:setting:model} briefly introduces the monomer interaction model used in our simulations.
Section \ref{subsec:setting:collision} describes our simulation setups, including the methods to prepare initial pebbles.

\subsection{Monomer Interaction Model}
\label{subsec:setting:model}

We calculate the interactions of contacting monomers as elastic spheres using a theoretical model of \citet{Dominik1997} and \citet{Wada2007} based on Johnson--Kendall--Roberts theory \citep{Johnson1971}.
This model provides monomer interactions associated with normal, sliding, rolling, and twisting motions.
These interactions are characterized by material parameters listed in Table \ref{tab:parameters}: monomer radius $r_0$, material density $\rho_0$, surface energy $\gamma$, Poisson's ratio $\nu$, Young's modulus $E$, and the critical rolling displacement $\xi_\mathrm{crit}$.

\begin{deluxetable}{lcc}
\tablecaption{Material Parameters of Ice and Silicate \label{tab:parameters}}
\tablehead{
\colhead{Parameter} & \colhead{Ice} & \colhead{Silicate}
}
\startdata 
Monomer radius $r_0$ ($\mathrm{\mu m}$) & 0.1 & 0.1 \\
Material density $\rho_0$ (g cm$^{-3}$) & 1.0 & 2.65 \\
Surface energy $\gamma$ (mJ m$^{-2}$) & 100 & 20 \\
Poisson's ratio $\nu$ & 0.25 & 0.17 \\
Young's modulus $E$ (GPa) & 7 & 54 \\
Critical rolling displacement $\xi_\mathrm{crit}$ (\AA) & 8 & 20 \\
\enddata
\tablecomments{We use $\gamma$ and $\xi_\mathrm{crit}$ of ice of \citet{Israelachvili1992} and \citet{Dominik1997} and of silicate of \citet{Seizinger2012}.}
\end{deluxetable}

Among the monomer interactions, rolling friction determines the compressive strength of dust aggregates \citep{Kataoka2013,Tatsuuma2023}.
Two contacting monomers roll irreversibly after the absolute value of the rolling displacement exceeds a threshold $\xi_\mathrm{crit}$, which has been estimated to be 2 {\AA} \citep{Dominik1997} and 32 {\AA} \citep{Heim1999} based on theoretical and experimental studies, respectively.
In this work, we adopt $\xi_\mathrm{crit}=8$ {\AA} for ice, as the geometric mean of the above two values, and $\xi_\mathrm{crit}=20$ {\AA} for silicate according to \citet{Seizinger2012}.
The energy needed for a monomer to roll a distance of $(\pi/2)r_0$ is given as
\begin{equation}
E_\mathrm{roll}=6\pi^2\gamma r_0\xi_\mathrm{crit}.
\end{equation}

To suppress relative oscillation between two contacting monomers, we add a damping force in the normal direction, which mimics the viscoelasticity or hysteresis of monomers \citep[e.g.,][]{Greenwood2006,Tanaka2012,Krijt2013}.
As described in Section 2.2 of \citet{Tatsuuma2019}, the damping force is proportional to a dimensionless coefficient $k_\mathrm{n}$.
In this work, we use $k_\mathrm{n}=0.1$ for initial pebble preparation and $k_\mathrm{n}=0$ for collision simulations so that the contact surfaces are not affected by the damping force.

\subsection{Simulation Setups}
\label{subsec:setting:collision}

\begin{figure*}[t!]
\plotone{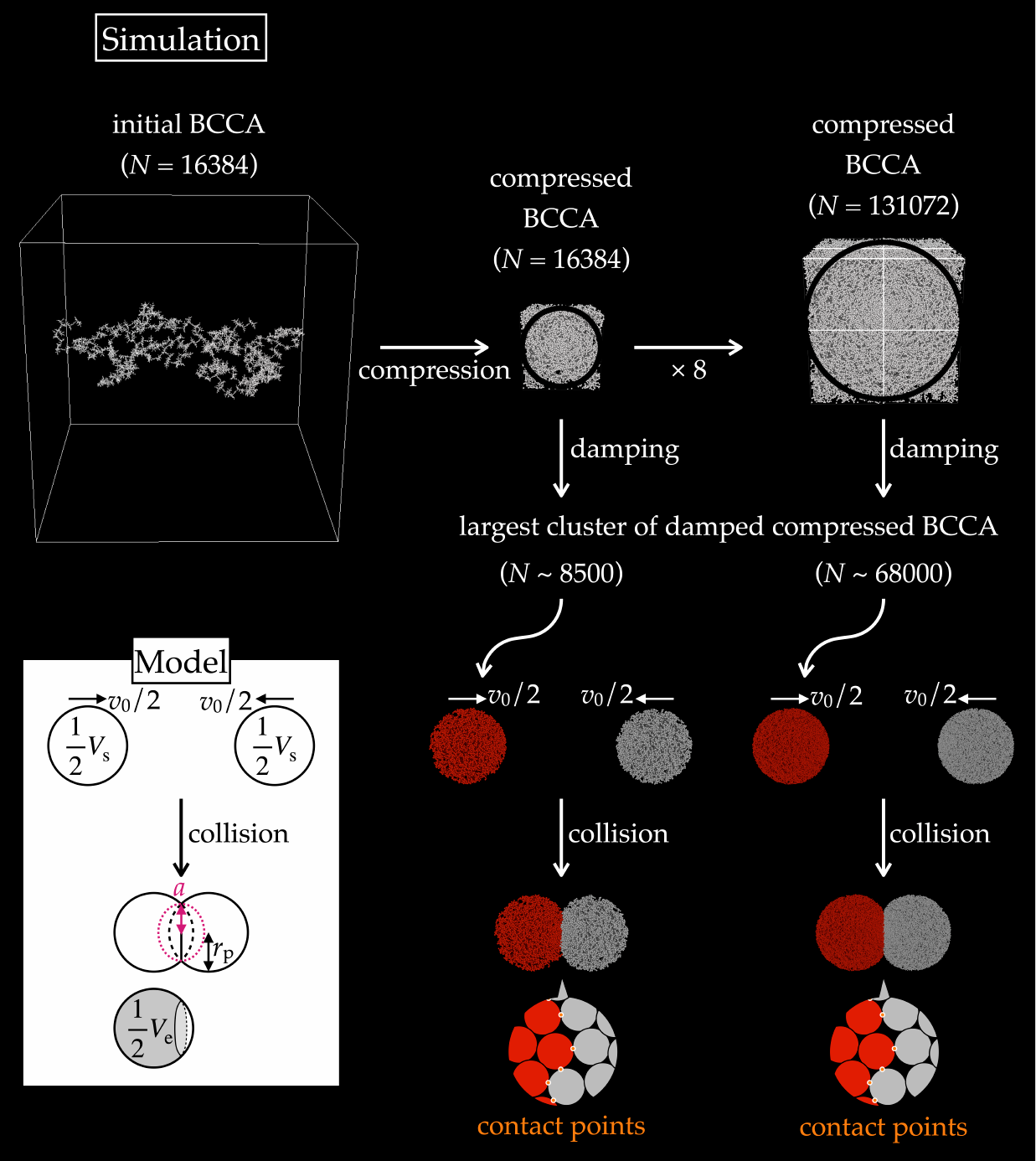} 
\caption{Schematic of collision simulations of pebbles.
We generate a BCCA and compress it in three dimensions using periodic boundaries.
We extract a spherical region from the compressed BCCA for the fiducial cases.
For larger aggregates, we combine eight compressed BCCAs before carving out a spherical region.
We suppress the monomer oscillations and remove isolated monomers.
We let two identical aggregates collide with random relative orientations.
After the collision at a velocity of $v_0$, contact points are determined as the midpoints of two contacting monomers from different pebbles.
\label{fig:outline}}
\end{figure*}

To prepare initial pebbles, i.e., initial dust aggregates, we follow the method described in Figure \ref{fig:outline}.
We generate a ballistic cluster--cluster aggregate (BCCA) composed of 16384 monomers and compress it in three dimensions using periodic boundaries \citep[e.g.,][]{Kataoka2013,Tatsuuma2023} to volume filling factors of $\phi_\mathrm{s}=0.1$, 0.2, 0.3, and 0.4.
This process simulates the natural coagulation and compression of dust aggregates.
Subsequently, for fiducial cases, we carve out a spherical region from the compressed BCCA.
Additionally, we construct a larger aggregate by merging eight identical compressed BCCAs before carving out a spherical region.
We suppress monomer oscillations using $k_\mathrm{n}=0.1$ and remove any monomers that are not connected to the main body.

In a collision simulation, we let two identical aggregates collide with random relative orientations.
The initial velocities of each pebble, $v_0/2=3$, 10, 30, 100, 300, and 1000 $\mathrm{cm\ s^{-1}}$, are given in opposite directions so that the pebbles collide.

The contact point is defined as the midpoint of two contacting monomers from different pebbles.
To exclude contacts involving isolated monomers, each contacting monomer must also be in contact with other monomers within the same pebble.

\section{Model of Contact Surface Radius}
\label{sec:model} 

Before presenting our simulation results, we develop an analytical model to predict the contact surface radius $a$ for a given relative collision velocity $v_0$.
In this model, we make three key assumptions regarding the volume of pebbles, the volume filling factor of the compressed region, and the conversion of kinetic energy.

We assume that pebbles are initially spherical and become truncated spheres after collision.
The initial volume of the two spheres, $V_\mathrm{s}$, and the final volume of the two truncated spheres, $V_\mathrm{e}$, as shown in Figure \ref{fig:outline}, are given as follows:
\begin{eqnarray}
V_\mathrm{s} &=& 2\times\frac{4}{3}\pi r_\mathrm{p}^3, \\
V_\mathrm{e} &=& 2\times\frac{\pi}{3}\left[(2r_\mathrm{p}^2+a^2)\sqrt{r_\mathrm{p}^2-a^2}+2r_\mathrm{p}^3\right],
\end{eqnarray}
where $r_\mathrm{p}$ is the pebble radius.
Therefore, the decreased volume is given as
\begin{eqnarray}
V_\mathrm{r} &=& V_\mathrm{s}-V_\mathrm{e}\nonumber\\
&=& \frac{2}{3}\pi r_\mathrm{p}^3\left\{2-\left[2+\left(\frac{a}{r_\mathrm{p}}\right)^2\right]\sqrt{1-\left(\frac{a}{r_\mathrm{p}}\right)^2}\right\}.
\label{eq:V_r}
\end{eqnarray}

{\reviseone{We assume that only a region close to the contact surface is uniformly compressed.
The volume filling factor in the compressed region increases from $\phi_\mathrm{s}$ to $\phi_\mathrm{e}$.
The initial volume of the compressed region is set to $Ca^3+V_\mathrm{r}$, and compression reduces this volume to $Ca^3$, where $C$ is a numerical constant.
Since the number of monomers in the compressed region remains constant, we obtain}}
\begin{equation}
(Ca^3+V_\mathrm{r})\phi_\mathrm{s} = Ca^3\phi_\mathrm{e},
\label{eq:volume-filling}
\end{equation}
and combining Equations (\ref{eq:V_r}) and (\ref{eq:volume-filling}) yields
\begin{eqnarray}
\phi_\mathrm{e} &=& \left\{1+\frac{4\pi}{3C}\left(\frac{a}{r_\mathrm{p}}\right)^{-3}\right.\nonumber\\
&&-\left.\frac{2\pi}{3C}\left[2\left(\frac{a}{r_\mathrm{p}}\right)^{-2}+1\right]\sqrt{\left(\frac{a}{r_\mathrm{p}}\right)^{-2}-1}\right\}\phi_\mathrm{s}.
\label{eq:phi_e}
\end{eqnarray}

Finally, we assume that the initial kinetic energy of the pebbles is entirely used for inelastic compression.
We adopt the relationship between compressive strength and volume filling factor derived by \citet{Tatsuuma2023}, who conducted numerical simulations of the three-dimensional compression of dust aggregates.
The compressive strength $P_\mathrm{comp}$ is given as
\begin{equation}
P_\mathrm{comp} = \frac{E_\mathrm{roll}}{r_0^3}\left(\frac{1}{\phi}-\frac{1}{\phi_\mathrm{max}}\right)^{-3},
\label{eq:comp}
\end{equation}
where $\phi$ is the volume filling factor of dust aggregates and $\phi_\mathrm{max}=\sqrt{2}\pi/6=0.74$ is the volume filling factor of the closest packing.
{\reviseone{Therefore, the equivalence between the initial kinetic energy and the work for the compression is expressed as}}
\begin{equation}
m_\mathrm{p}\left(\frac{v_0}{2}\right)^2 = -\int_{Ca^3+V_\mathrm{r}}^{Ca^3}\frac{E_\mathrm{roll}}{r_0^3}\left(\frac{1}{\phi}-\frac{1}{\phi_\mathrm{max}}\right)^{-3}dV,
\end{equation}
where $m_\mathrm{p}$ is the pebble mass.

We convert the volume integral into a density integral using
\begin{eqnarray}
\rho &=& \frac{Nm_0}{V}
= \rho_0\phi,\\
dV &=& -\frac{Nm_0}{\rho^2}d\rho,
\end{eqnarray}
where $\rho$ and $V$ represent the density and volume of the compressed region, respectively, and $m_0=(4/3)\pi r_0^3\rho_0$ denotes the monomer mass.
We obtain
\begin{eqnarray}
m_\mathrm{p}\left(\frac{v_0}{2}\right)^2 &=& -\int_{Ca^3+V_\mathrm{r}}^{Ca^3}\frac{E_\mathrm{roll}}{r_0^3}\left(\frac{1}{\phi}-\frac{1}{\phi_\mathrm{max}}\right)^{-3}dV \nonumber \\
&=& \frac{Nm_0E_\mathrm{roll}}{r_0^3}\int_{\rho_\mathrm{s}}^{\rho_\mathrm{e}}\left(\frac{\rho_0}{\rho}-\frac{1}{\phi_\mathrm{max}}\right)^{-3}\frac{d\rho}{\rho^2},
\end{eqnarray}
where the density increases from $\rho_\mathrm{s}=\phi_\mathrm{s}\rho_0$ to $\rho_\mathrm{e}=\phi_\mathrm{e}\rho_0$.
Rewriting $N$ as
\begin{equation}
N = \frac{\rho V}{m_0}
= \frac{\rho_\mathrm{e}Ca^3}{m_0}
= \frac{\phi_\mathrm{e}\rho_0Ca^3}{m_0}
= \frac{\phi_\mathrm{s}\rho_0(Ca^3+V_\mathrm{r})}{m_0},
\end{equation}
we obtain
\begin{eqnarray}
v_0^2
&=& {\frac{2CE_\mathrm{roll}}{m_0}}\frac{\phi_\mathrm{e}}{\phi_\mathrm{s}}\left(\frac{a}{r_\mathrm{p}}\right)^3\nonumber\\
&&\times\left[\left(\frac{1}{\phi_\mathrm{e}}-\frac{1}{\phi_\mathrm{max}}\right)^{-2}-\left(\frac{1}{\phi_\mathrm{s}}-\frac{1}{\phi_\mathrm{max}}\right)^{-2}\right].
\label{eq:v_0}
\end{eqnarray}
To determine $a/r_\mathrm{p}$ for a given $v_0$, {\reviseone{it is important to note}} that $\phi_\mathrm{e}$ depends on $a/r_\mathrm{p}$, as shown in Equation (\ref{eq:phi_e}).
Thus, by numerically solving Equations (\ref{eq:phi_e}) and (\ref{eq:v_0}) for $a/r_\mathrm{p}$ with respect to the given $v_0$, we obtain $a/r_\mathrm{p}$ as a function of $v_0$.
In Section \ref{sec:result}, we validate Equation (\ref{eq:v_0}) and determine the value of $C$ based on our simulation results.

We also considered an alternative model in which the {\reviseone{volume of the compressed region}} is assumed to decrease from $C_2a^2\delta+V_\mathrm{r}$ to $C_2a^2\delta$, where $C_2$ is a constant and $\delta=r_\mathrm{p}-\sqrt{r_\mathrm{p}^2-a^2}$ is the thickness of the compressed region.
However, we find that this alternative model provides a less accurate fit to the simulation data (see Appendix \ref{apsec:anothermodel}).

\section{Comparison of Model and Simulations} \label{sec:result}

We present our simulation results to determine the contact surface radius and the number of contact points on the contact surface in Sections \ref{subsec:result:contactsurface} and \ref{subsec:result:N_con}, respectively.

\subsection{Contact Surface Radius}
\label{subsec:result:contactsurface}

In this section, we present the results for the contact surface radius.
Section \ref{subsubsec:result:contactsurface:measure} describes the measurement of the contact surface radius using the characteristic radius.
We compare the simulation results with our model in Sections \ref{subsubsec:result:contactsurface:ice} and \ref{subsubsec:result:contactsurface:silicate} for ice and silicate pebbles, respectively.

\subsubsection{Measuring Contact Surface Radius of Simulated Aggregates}
\label{subsubsec:result:contactsurface:measure}

To measure the contact surface radius $a$, we approximate the discrete points of the contact surface as a circle using the characteristic radius $a_\mathrm{c,2d}$ \citep[e.g.,][]{Mukai1992,Okuzumi2009dustagg}.

The characteristic radius is derived in both two and three dimensions using the radius of gyration defined by
\begin{equation}
a_\mathrm{g} \equiv \sqrt{\frac{1}{N}\sum^N_{k=1}|\bm{x}_k-\bm{X}|^2},
\end{equation}
where $\bm{x}_k$ is the $k$-th particle position vector, $\bm{X}$ is the geometric center of the contact region or sphere, and $N$ is the number of particles.
Here, we set $\bm{X}$ as the origin when calculating the contact surface radius, and as the geometric center when calculating the pebble radius.

To derive the contact surface radius, we assume a uniform number density $n$ inside a circle $S$ with a radius of $a_\mathrm{c,2d}$, and express the radius of gyration as
\begin{eqnarray}
a_\mathrm{g}^2 &=& \frac{\sum^N_{k=1}|\bm{x}_k-\bm{X}|^2}{N} 
= \frac{\int_Sn r^2 dS}{\int_Sn dS}
= \frac{\int_0^{a_\mathrm{c,2d}} r^2 2\pi r dr}{\pi a_\mathrm{c,2d}^2} \nonumber \\
&=& \frac{1}{2}a_\mathrm{c,2d}^2.
\label{eq:gyration2d}
\end{eqnarray}
Equation (\ref{eq:gyration2d}) gives
\begin{equation}
a \equiv a_\mathrm{c,2d} = \sqrt{2}a_\mathrm{g}.
\label{eq:chara2d}
\end{equation}

Similarly, the pebble radius $r_\mathrm{p}$ is derived by assuming a sphere $V$ with a radius of $a_\mathrm{c,3d}$ and expressing the radius of gyration as
\begin{eqnarray}
a_\mathrm{g}^2 &=& \frac{\sum^N_{k=1}|\bm{x}_k-\bm{X}|^2}{N} 
= \frac{\int_Vn r^2 dV}{\int_Vn dV}
= \frac{\int_0^{a_\mathrm{c,3d}} r^2 4\pi r^2 dr}{(4/3)\pi a_\mathrm{c,3d}^3} \nonumber \\
&=& \frac{3}{5}a_\mathrm{c,3d}^2,
\label{eq:gyration3d}
\end{eqnarray}
which gives
\begin{equation}
r_\mathrm{p} \equiv a_\mathrm{c,3d} = \sqrt{\frac{5}{3}}a_\mathrm{g}.
\label{eq:chara3d}
\end{equation}

\begin{figure*}[t!]
\plotone{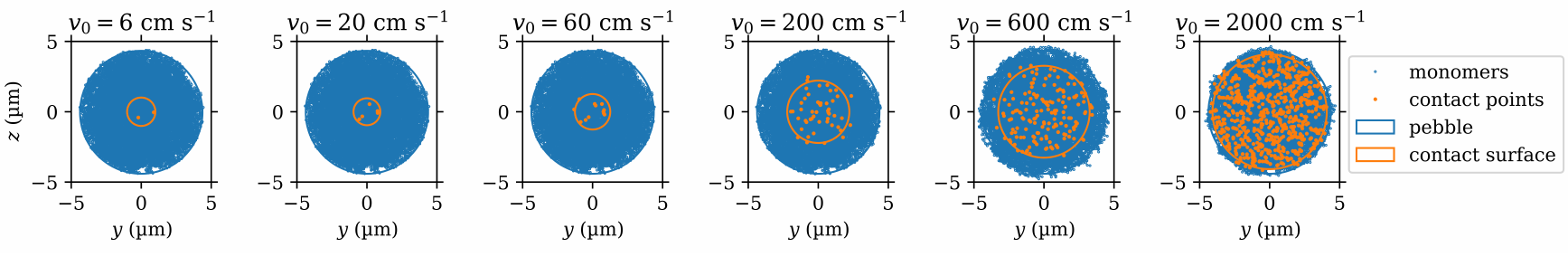} 
\caption{Projected $y$-$z$ plane coordinates of monomers (blue) and contact points (orange) of two colliding pebbles along the $x$-axis.
Each pebble has $N_\mathrm{p}=8555$ monomers, with an initial volume filling factor of $\phi_\mathrm{s}=0.1$.
The two circles represent the pebble radius (blue, from Equation (\ref{eq:chara3d})) and the contact surface radius (orange, from Equation (\ref{eq:chara2d})).
\label{fig:contact_area_radius_phi_01}}
\end{figure*}

Figure \ref{fig:contact_area_radius_phi_01} shows the contact surface area for fiducial runs of ice pebbles with $N_\mathrm{p}=8555$ and $\phi_\mathrm{s}=0.1$ after collisions at different velocities of $v_0=6$, 20, 60, 200, 600, and 2000 $\mathrm{cm\ s^{-1}}$, where $N_\mathrm{p}$ is the number of monomers for a pebble.
The pebbles collide along the $x$-axis, and the coordinates of their monomers and contact points are projected onto the $y$-$z$ plane.
Figure \ref{fig:contact_area_radius_phi_01} also shows the pebble radius $r_\mathrm{p}\simeq4.4\mathrm{\ \mu m}$ from Equation (\ref{eq:chara3d}) and the contact surface radius $a$ from Equation (\ref{eq:chara2d}).
The number of contact points increases with $v_0$, forming a well-defined contact area at $v_0 > 6 \mathrm{\ cm\ s^{-1}}$.
For such high velocities, $a_\mathrm{c,2d}$, as defined in Equation (\ref{eq:chara2d}), accurately represents the contact surface radius.

To estimate the error of $a_\mathrm{c,2d}$, we account for the error arising from the limited number of contact points, instead of using the variance of collision simulations by varying the initial relative orientations of the pebbles, which would lead to an unrealistically small error estimate.
We conduct the Monte Carlo calculations in Appendix \ref{apsec:errorMC} and use Equation (\ref{eq:error}) for error estimation.

\subsubsection{Fitting and Parameter Dependencies for Ice Pebbles}
\label{subsubsec:result:contactsurface:ice}

\begin{figure*}[t!]
\plotone{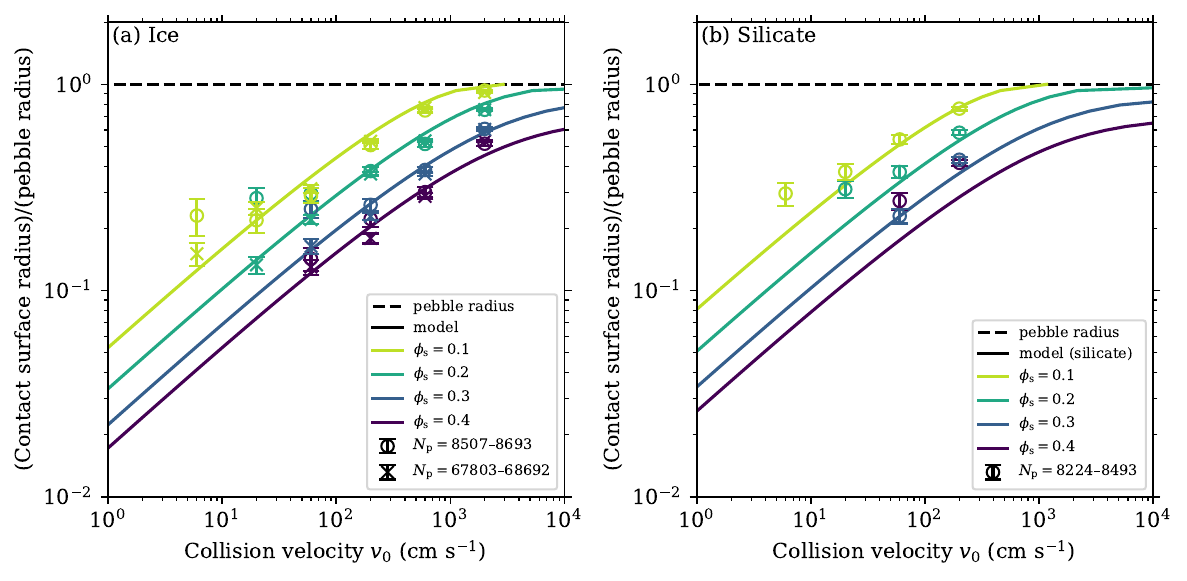} 
\caption{Left: relationship between the collision velocity $v_0$ and $a/r_\mathrm{p}$ for ice pebbles.
The dashed and solid lines represent the pebble radius and our model (Equations (\ref{eq:phi_e}) and (\ref{eq:v_0})), respectively.
The circles and crosses are simulation results with $N_\mathrm{p}=8507$--8693 and 67803--68692, respectively.
The initial volume filling factors are $\phi_\mathrm{s}=0.1$, 0.2, 0.3, and 0.4 from light to dark colors.
The error is estimated with the Monte Carlo method as described in Appendix \ref{apsec:errorMC}.
Right: relationship between the collision velocity $v_0$ and $a/r_\mathrm{p}$ for silicate pebbles.
\label{fig:velocity_contact_area}}
\end{figure*}

Figure \ref{fig:velocity_contact_area}(a) shows the relationship between the collision velocity $v_0$ and $a/r_\mathrm{p}$ from our simulation results for ice pebbles with different initial volume filling factors, $\phi_\mathrm{s}$, and the number of monomers per pebble, $N_\mathrm{p}$.

We fit the constant $C$ in our model (Equations (\ref{eq:phi_e}) and (\ref{eq:v_0})) to the simulation results and then obtain $C=1.5$; $C\approx1$ as expected.
Here, we have used the simulation data with $v_0=200$--2000 $\mathrm{cm\ s^{-1}}$ for smaller $N_\mathrm{p}=8507$--8693 and all simulation data for larger $N_\mathrm{p}=67803$--68692, because the number of contact points is approximately 1 when $N_\mathrm{p}$ and $v_0$ are too small, which makes the continuous model invalid.
The details of the fitting are described in Appendix \ref{apsec:fitting}.

The dependencies of the simulation results and the model predictions on each parameter are as follows.
The observed dependence on the collision velocity, $v_0$, is consistent with our model.
According to our model, the relationship between $v_0$ and $a/r_\mathrm{p}$ for low collision velocities follows the power law, $a/r_\mathrm{p}\propto v_0^{1/2}$, as derived in Appendix \ref{apsec:modelinterp}.
The dependence on the initial volume filling factor $\phi_\mathrm{s}$ is well described by our model because Equation (\ref{eq:comp}) describes the effect of $\phi_\mathrm{max}$, which becomes significant when $\phi>0.1$ \citep{Tatsuuma2023}.
Our model predicts that for a fixed collision velocity, $v_0$, $a$ scales linearly with $r_\mathrm{p}$, which is confirmed by the simulation results.
For smaller $N_\mathrm{p}$ and lower $v_0$, the contact surfaces cannot be resolved by monomers, leading to an overestimation in the simulations.

\subsubsection{Parameter Dependencies for Silicate Pebbles}
\label{subsubsec:result:contactsurface:silicate}

We find that $C=1.5$ also applies to silicate pebbles with $v_0=6$--200 $\mathrm{cm\ s^{-1}}$, as shown in Figure \ref{fig:velocity_contact_area}(b).

The dependence on the initial volume filling factor $\phi_\mathrm{s}$ is well described by our model when $\phi_\mathrm{s}\leq0.3$, while our model underestimates the contact surface radius when $\phi_\mathrm{s}=0.4$.
{\reviseone{In other words, silicate pebbles deform more than expected for high volume filling factors.}}

\begin{figure*}[t!]
\plotone{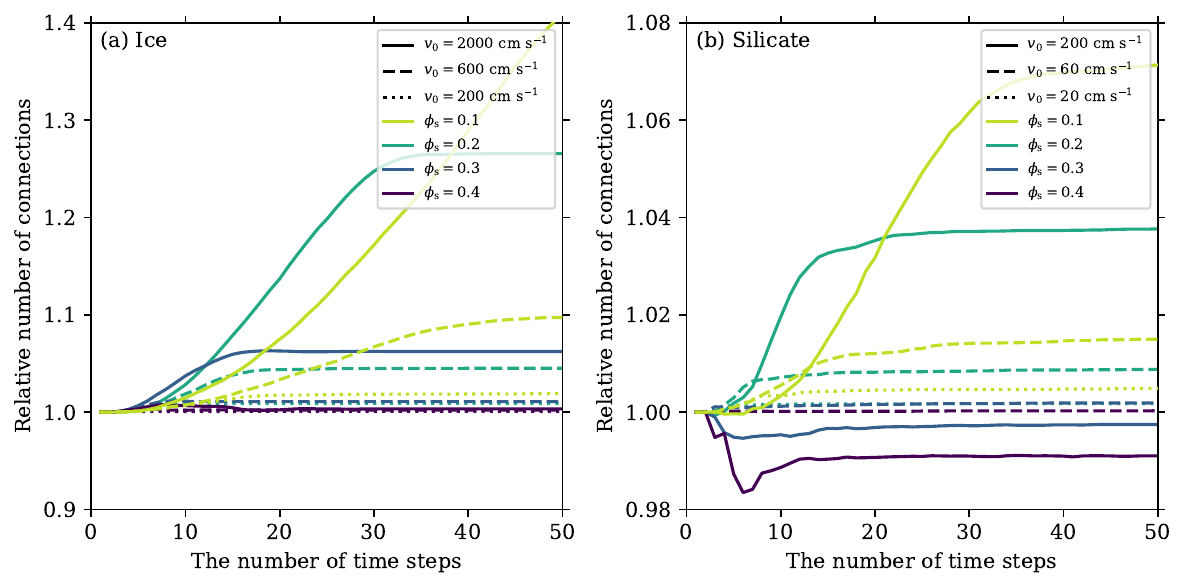} 
\caption{Left: relative number of connections between monomers against the number of time steps for ice pebbles.
The solid, dashed, and dotted lines represent $v_0=2000$, 600, and 200 $\mathrm{cm\ s^{-1}}$.
The initial volume filling factors are $\phi_\mathrm{s}=0.1$, 0.2, 0.3, and 0.4 from light to dark colors.
Right: relative number of connections between monomers against the number of time steps for silicate pebbles.
The solid, dashed, and dotted lines represent $v_0=200$, 60, and 20 $\mathrm{cm\ s^{-1}}$.
\label{fig:connectnum}}
\end{figure*}

One possible reason is that a high volume filling factor may lead to internal fragmentation rather than deformation due to rolling, twisting, and sliding motions.
This means that silicate pebbles are fragile, as indicated by their lower critical fragmentation velocity compared to ice due to the lower surface energy of silicate \citep[e.g.,][]{Wada2013}.
Figure \ref{fig:connectnum} shows that the number of connections between monomers decreases for silicate pebbles when $(\phi_\mathrm{s},v_0)=(0.3,200\mathrm{\ cm\ s^{-1}})$ and $(0.4,200\mathrm{\ cm\ s^{-1}})$.
In contrast, the number of connections increases in all other cases.

\subsection{The Number of Contact Points}
\label{subsec:result:N_con}

The number of contact points on the contact surface is a key factor in determining the mechanical strength and thermal conductivity of a pebble aggregate.
In this section, we derive an analytical expression for predicting the number of contact points as a function of the contact surface radius.

\begin{figure}[t!]
\plotone{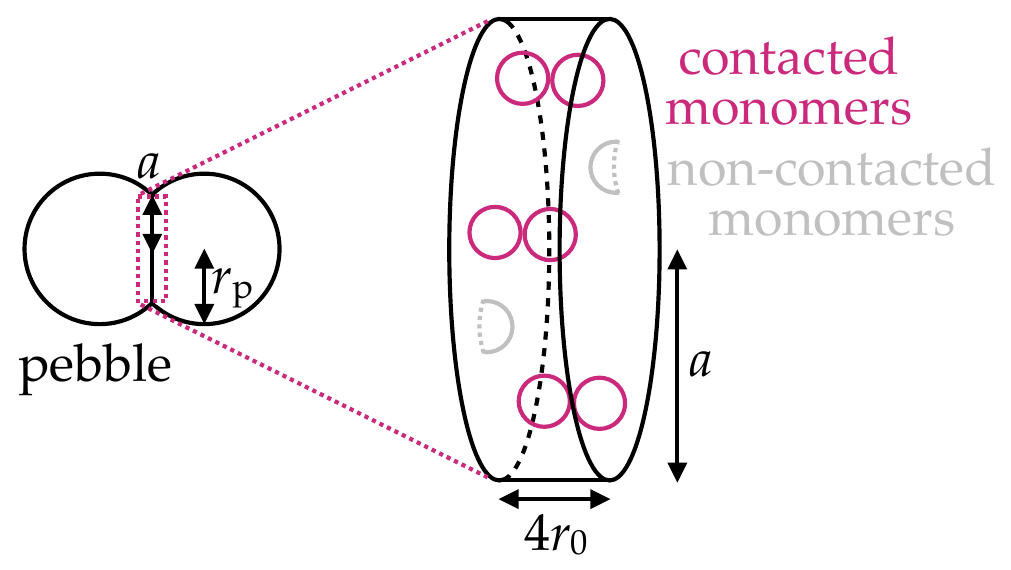}
\caption{Schematic of the cylindrical region near the contact surface.
It contains pairs of monomers that are in contact and contribute to contact points, as well as monomers that do not contribute to pebble contact.
\label{fig:nearcontact}}
\end{figure}

To estimate the number of contact points on the contact surface, $N_\mathrm{con}$, we consider a cylindrical region near the contact surface with a base radius $a$ and a thickness of $4r_0$, as shown in Figure \ref{fig:nearcontact}.
We approximate the volume filling factor in this region as $\phi_\mathrm{e}$ given by Equation (\ref{eq:phi_e}).
The upper limit of $N_\mathrm{con}$ is estimated as
\begin{equation}
N_\mathrm{con} \leq \frac{\pi a^2 \cdot 4r_0 \cdot \phi_\mathrm{e}}{(4/3)\pi r_0^3}\cdot\frac{1}{2}
= \frac{3}{2}\phi_\mathrm{e}\frac{a^2}{r_0^2}.
\end{equation}
Since this volume region also contains monomers that do not contribute to pebble contact, we introduce a constant $C_\mathrm{con}\lesssim1$ and obtain the number of contact points given by
\begin{equation}
N_\mathrm{con} = C_\mathrm{con} \phi_\mathrm{e}\frac{a^2}{r_0^2}.\label{eq:N_con}
\end{equation}

\begin{figure*}[t!]
\plotone{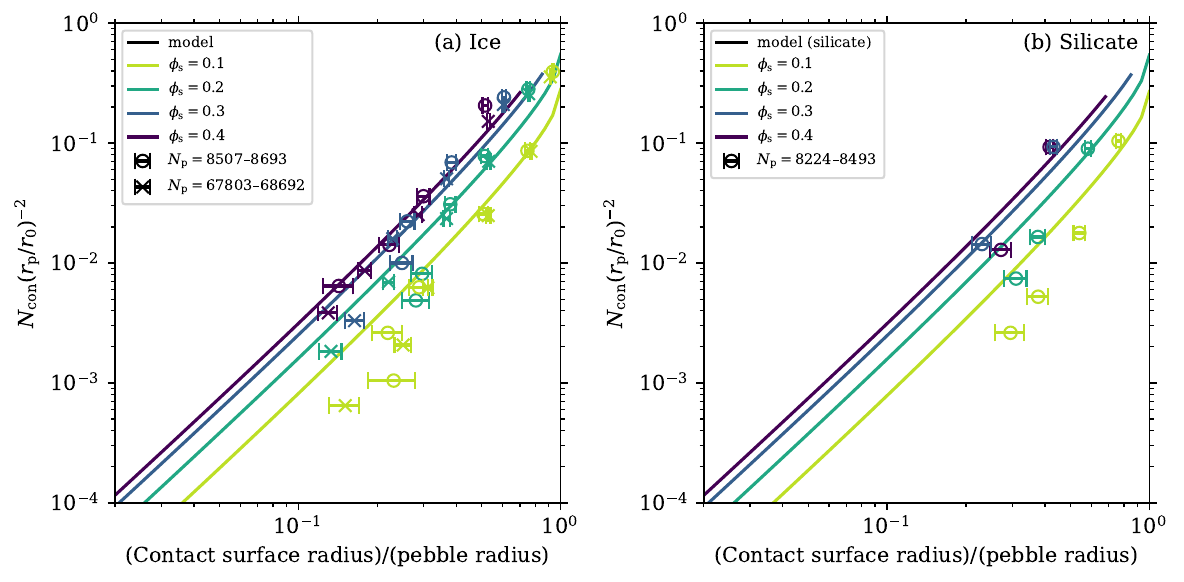} 
\caption{Left: relationship between $a/r_\mathrm{p}$ and $N_\mathrm{con}(r_\mathrm{p}/r_0)^{-2}$ for ice pebbles.
The solid lines represent our model (Equations (\ref{eq:phi_e}) and (\ref{eq:N_con})).
The other lines, markers, and colors are the same as in Figure \ref{fig:velocity_contact_area}.
Right: relationship between $a/r_\mathrm{p}$ and $N_\mathrm{con}(r_\mathrm{p}/r_0)^{-2}$ for silicate pebbles.
\label{fig:N_con}}
\end{figure*}

Figure \ref{fig:N_con} shows $N_\mathrm{con}$ as a function of $a$ for colliding pebbles measured in our simulations.
Here, $N_\mathrm{con}$ and $a$ are normalized by $(r_\mathrm{p}/r_0)^{2}$ and $r_\mathrm{p}$, respectively, because Equation (\ref{eq:N_con}) implies that $N_\mathrm{con}(r_\mathrm{p}/r_0)^{-2}$ is determined solely by $(a/r_\mathrm{p})$.
By fitting Equation (\ref{eq:N_con}) with $\phi_\mathrm{e}$ given by Equation (\ref{eq:phi_e}) to the simulation results, we obtain $C_\mathrm{con} = 0.72$.
The details of the fitting procedures are described in Appendix \ref{apsec:fitting}.

\section{Application to Comet 67P}
\label{sec:apply}

{\revisetwo{In this section, we apply the pebble contact model developed in this work to solid bodies composed of pebbles.
The model relates the contact surface radius between pebbles to their collision velocity.
By estimating the contact surface radius from either the tensile strength or thermal conductivity of the body, we can infer the collision velocity between constituent pebbles.}}
The inferred velocity places constraints on the formation process of the body.

Here, we apply this analysis to the well-studied comet 67P, whose tensile strength has been estimated.
Analyses of surface features on 67P by \citet{Groussin2015} suggest very low strengths: $\sim3$--15 Pa for small overhangs ($\sim10$ m) and $<150$ Pa for larger collapsed structures ($\sim100$ m).
\citet{Attree2018} found even smaller values, typically 0--5 Pa, corresponding to a bulk tensile strength of about 1 Pa at the decametre scale.
Other constraints from the breakup of sungrazing and rotating comets indicate somewhat higher values, up to $\sim100$ Pa \citep{Klinger1989,Davidsson2001}.
Similarly, \citet{Basilevsky2016} reported a range of 1.5--100 Pa, while \citet{Hirabayashi2016} suggested 10--200 Pa.
Taking these estimates together, we adopt a representative range of $T_\mathrm{agg}=1.5$--100 Pa for the tensile strength of 67P.

Its bulk density has also been estimated to be $\rho_\mathrm{bulk}=0.532 \mathrm{\ g\ cm^{-3}}$ from measurements of its volume and mass \citep{Jorda2016}.
In Section \ref{subsec:apply:contactradius}, we estimate the contact surface radius between pebbles from the tensile strength of 67P.
Subsequently, in Section \ref{subsec:apply:velocity}, the collision velocity between pebbles is estimated from the obtained contact surface radius.
Finally, in Section \ref{subsec:apply:bounce}, we discuss possible mechanisms that could explain the obtained collision velocity.

\subsection{Estimating Contact Radius Ratio}
\label{subsec:apply:contactradius}

We estimate $a/r_\mathrm{p}$ from the tensile strength of the comet as a pebble aggregate, $T_\mathrm{agg}$, and that for individual pebbles, considered as dust aggregates, $T_\mathrm{p}$.
The relationship between $T_\mathrm{agg}$ and $T_\mathrm{p}$ is given by
\begin{equation}
T_\mathrm{agg} \pi r_\mathrm{p}^2 = T_\mathrm{p} \pi a^2,
\end{equation}
and $a/r_\mathrm{p}$ is obtained as
\begin{equation}
\frac{a}{r_\mathrm{p}}=\sqrt{\frac{T_\mathrm{agg}}{T_\mathrm{p}}}.
\label{eq:a_rp_Tagg_Tp}
\end{equation}
The tensile strength of dust aggregates, $T_\mathrm{p}$, is estimated as \citep{Tatsuuma2019}
\begin{equation}
T_\mathrm{p} = 6\times 10^5 \mathrm{\ Pa} \left(\frac{\gamma}{100\mathrm{\ mJ\ m^{-2}}}\right)\left(\frac{r_0}{0.1\mathrm{\ \mu m}}\right)^{-1}\phi_\mathrm{p}^{1.8},
\label{eq:tensiledustagg}
\end{equation}
where $\phi_\mathrm{p}$ is the microscopic filling factor of individual pebbles.
{\reviseone{For given $T_\mathrm{agg}=1.5$--100 Pa, Equation (\ref{eq:a_rp_Tagg_Tp}) gives $a/r_\mathrm{p}$ as a function of $\gamma$, $r_0$, and $\phi_\mathrm{p}$.}}

This microscopic filling factor should be distinguished from the bulk volume filling factor, which is given as
\begin{equation}
\phi_\mathrm{bulk}=\frac{\rho_\mathrm{bulk}}{\rho_0}=\phi_\mathrm{p}\phi_\mathrm{agg},
\end{equation}
where $\phi_\mathrm{agg}$ is the macroscopic filling factor of the comet as a pebble aggregate.
Assuming monodisperse pebbles, $\phi_\mathrm{agg}$ has an upper limit $\phi_\mathrm{max}=0.74$ at closest packing.
Therefore, the microscopic filling factor $\phi_\mathrm{p}$ must have a range given as
\begin{equation}
\frac{\rho_\mathrm{bulk}}{\rho_0\phi_\mathrm{max}} < \frac{\rho_\mathrm{bulk}}{\rho_0\phi_\mathrm{agg}} = \phi_\mathrm{p}
< {\phi_\mathrm{max}}.\label{eq:phi_pebble}
\end{equation}
We assume that 67P consists of only ice pebbles or only silicate pebbles.
For ice pebbles, we assume that each monomer consists of a silicate core and an ice mantle of equal mass.
The average material density is then $\rho_\mathrm{0,ice+sil}=1.45\mathrm{\ g\ cm^{-3}}$ \citep[Equation (5) of][]{Tatsuuma2024}.
From the estimated bulk density of 67P, $\rho_\mathrm{bulk}=0.532 \mathrm{\ g\ cm^{-3}}$, Equation (\ref{eq:phi_pebble}) gives 
\begin{equation}
{\phi_\mathrm{max}}>\phi_\mathrm{p} >
\left\{
\begin{array}{ll}
0.50 & \mathrm{(for\ ice)} \\
0.27 & \mathrm{(for\ silicate)},
\end{array}
\right.
\label{eq:phi_p}
\end{equation}
for ice and silicate pebbles, respectively.

\begin{figure}[t!]
\plotone{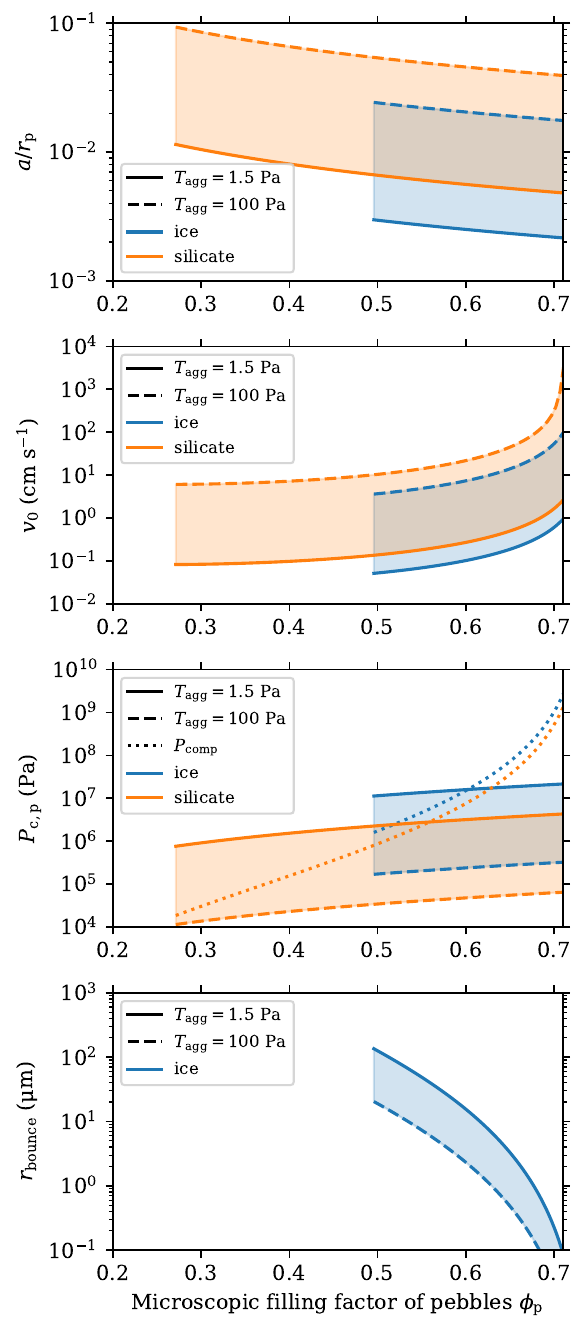}
\caption{Relationship between the microscopic filling factor of pebbles, $\phi_\mathrm{p}$ (common $x$-axis), and (from top to bottom): $a/r_\mathrm{p}$, $v_0$, $P_\mathrm{c,p}$, and $r_\mathrm{bounce}$.
The solid, dashed, and dotted lines represent $T_\mathrm{agg}=1.5$ Pa, 100 Pa, and $P_\mathrm{comp}$, respectively.
The blue and orange lines represent ice and silicate pebbles, respectively.
\label{fig:phi_p_depend}}
\end{figure}

Finally, we estimate $a/r_\mathrm{p}$ from the estimated tensile strength of 67P, $T_\mathrm{agg}=1.5$--100 Pa, using Equations (\ref{eq:a_rp_Tagg_Tp}), (\ref{eq:tensiledustagg}), and (\ref{eq:phi_p}).
Assuming $r_0=0.1\mathrm{\ \mu m}$, we obtain the relation between $\phi_\mathrm{p}$ and $a/r_\mathrm{p}$.
{\revisethree{The top panel of Figure \ref{fig:phi_p_depend} plots $a/r_\mathrm{p}$ as a function of $\phi_\mathrm{p}$ for different values of $T_\mathrm{agg}$ and composition.
The plot shows}}
\begin{equation}
\frac{a}{r_\mathrm{p}} \simeq
\left\{
\begin{array}{ll}
0.002\textrm{--}0.02 & \mathrm{(for\ ice)} \\
0.005\textrm{--}0.09 & \mathrm{(for\ silicate)},
\end{array}
\right.
\label{eq:67Pa_rp}
\end{equation}
for ice and silicate pebbles, respectively.

\subsection{Estimating Pebble Collision Velocity}
\label{subsec:apply:velocity}

Assuming that pebble collisions determine the contact surfaces and using Equations (\ref{eq:phi_e}), (\ref{eq:v_0}), and (\ref{eq:phi_p}), we estimate $v_0$ as a function of $\phi_\mathrm{p}$ from $a/r_\mathrm{p}$ obtained above as shown in the second panel of Figure \ref{fig:phi_p_depend}.
{\revisethree{We limit $\phi_\mathrm{p}$ to $\phi_\mathrm{p}\leq0.7$ as $v_0$ diverges as $\phi_\mathrm{p}\to0.74$.}}
The second panel of Figure \ref{fig:phi_p_depend} suggests that 67P formed via pebble collisions at low collision velocities below $\sim10$ cm s$^{-1}$, irrespective of the composition of the constituent pebbles, as long as $\phi_\mathrm{p}<0.6$.

{\reviseone{If we assume a larger monomer radius, the tensile strength of dust aggregates decreases, so the estimated $a/r_\mathrm{p}$ ($\propto r_0^{1/2}$) increases (Equations (\ref{eq:a_rp_Tagg_Tp}) and (\ref{eq:tensiledustagg})).
The collision velocity $v_0$ required to achieve this larger $a/r_\mathrm{p}$, however, is insensitive to $r_0$ (for low velocities, $v_0\propto (a/r_\mathrm{p})^2r_0^{-1}\propto r_0 r_0^{-1}$) since the compressive strength also decreases.}}

The above estimates assume that the compression of 67P by its own gravity has little effect on the pebble contact surfaces.
To assess this assumption, we estimate the central pressure of 67P and compare it with the static compressive strength of the pebbles.
{\reviseone{We estimate the central pressure of 67P, $P_\mathrm{c}$, from its diameter $D$ and bulk density $\rho_\mathrm{bulk}$ from the simple calculation given by
\begin{equation}
P_\mathrm{c} = G\frac{(M/2)^2}{\pi(D/2)^4}\simeq 100 \mathrm{\ Pa},
\end{equation}
where $M=4\pi(D/2)^3\rho_\mathrm{bulk}/3$ is the mass of 67P and $G$ is the gravitational constant.}}
Taking into account stress concentration effects, the pressure acting on individual pebbles, $P_\mathrm{c,p}$, is given by (shown in the third panel of Figure \ref{fig:phi_p_depend})
\begin{equation}
P_\mathrm{c,p} = P_\mathrm{c}\left(\frac{r_\mathrm{p}}{a}\right)^2.
\label{eq:P_c,p}
\end{equation}
We also plot the static compressive strength of dust aggregates given by Equations (\ref{eq:comp}) and (\ref{eq:phi_p}).

The third panel of Figure \ref{fig:phi_p_depend} shows the comparison between $P_\mathrm{c,p}$ and $P_\mathrm{comp}$.
{\revisethree{For $\phi_\mathrm{p}\gtrsim0.6$, $P_\mathrm{comp}$ exceeds $P_\mathrm{c,p}$, and thus self-gravity compression can be neglected.
For $\phi_\mathrm{p}\lesssim 0.6$, the outcome depends on the tensile strength of 67P.
If the tensile strength is 100 Pa, self-gravity can be ignored, whereas for 1.5 Pa, self-gravity becomes non-negligible because $P_\mathrm{c,p} > P_\mathrm{comp}$.
Even if self-gravity compression determines the contact surface between pebbles in the central region, its effect would be negligible in the near-surface layers that determine the tensile strength of 67P.
For the pebbles composing these latter regions, the conclusions of this work are applicable.}}

\subsection{Inelastic Bouncing as Origin of Low Collision Velocities}
\label{subsec:apply:bounce}

{\reviseone{The low pebble collision velocity inferred from the second panel of Figure \ref{fig:phi_p_depend} suggests that certain processes significantly damped the random motion of the pebbles that formed 67P.
Here, we explore the possibility that the pebbles lost their kinetic energy through repeated bouncing collisions.
It is known that relatively compact aggregates bounce upon colliding at sufficiently high velocities \citep[e.g.,][]{Kothe2013,Arakawa2023,Oshiro2025}.
These bouncing collisions are generally inelastic due to plastic deformation dissipating kinetic energy.
In the following, we test this scenario using the simulation results on aggregate collisions by \citet{Oshiro2025}, who conducted numerical simulations similar to this work, but with larger aggregate radii and higher volume filling factors.}}

\begin{figure*}[t!]
\plotone{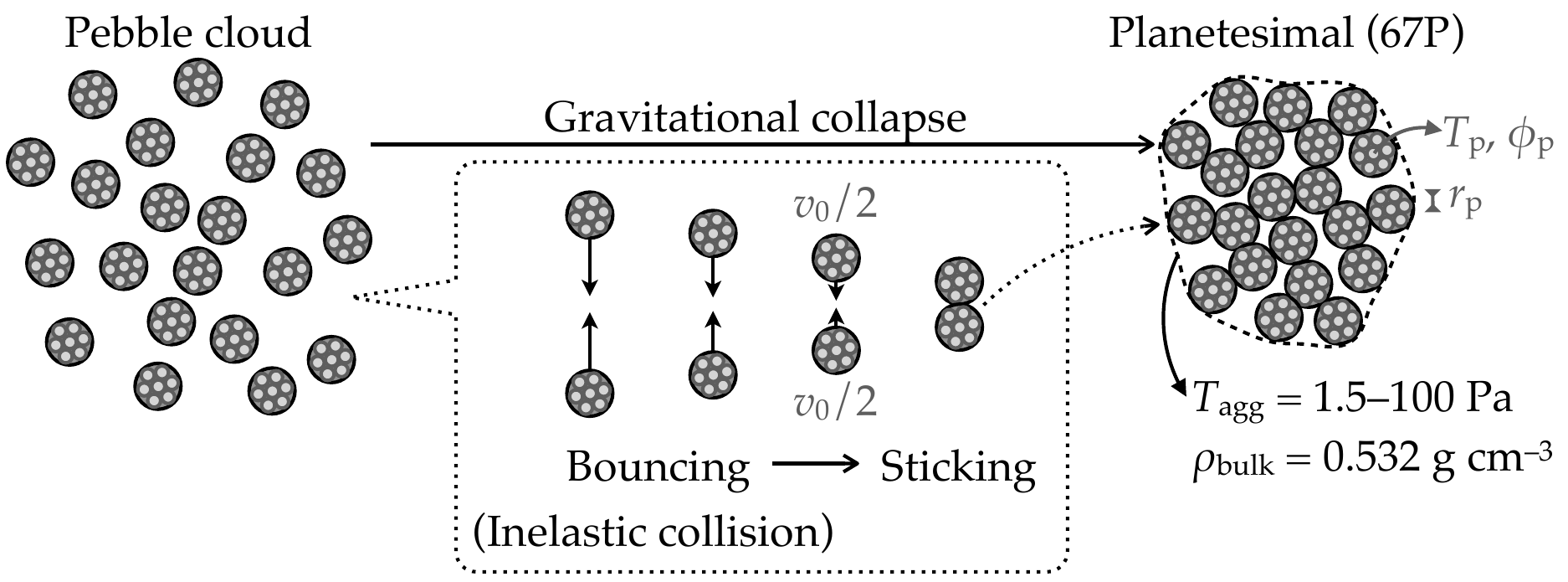} 
\caption{Schematic of the formation process of 67P.
A pebble cloud gravitationally collapses into a planetesimal.
Inside the collapsing cloud, pebbles sequentially bounce, lose their kinetic energy in inelastic collisions, and finally stick.
\label{fig:67Pformation}}
\end{figure*}

{\revisetwo{Here, we consider a scenario where 67P formed by the collapse of a gravitationally bound pebble cloud as illustrated in Figure \ref{fig:67Pformation}.
In this scenario, the 67P-forming pebbles inevitably undergo repeated inelastic collisions until their collision velocity becomes low enough to stick together.}}

{\reviseone{Below, we estimate the pebble size inside 67P from the bouncing scenario.
Previous studies of bouncing suggest that, when the collision velocity of bouncing pebbles falls below a critical value, sticking dominates over bouncing.
Also, this threshold velocity depends on the mass and compressive strength of pebbles.
Applying this scenario to the formation process of 67P, the collision velocity $v_0$ at the final sticking collisions should be equal to the threshold velocity for bouncing.
We use this hypothesis to predict the size of the pebbles that formed 67P.}}
The threshold mass for bouncing is given by Equations (11) and (17) of \citet{Oshiro2025} as
\begin{equation}
m_\mathrm{bounce}=2.4\times10^{-7}\mathrm{\ g}\left(\frac{P_\mathrm{comp}}{10^5\mathrm{\ Pa}}\right)^{-2.4}\left(\frac{v_0}{100\mathrm{\ cm\ s^{-1}}}\right)^{-4/3}.
\label{eq:m_bounce}
\end{equation}
The corresponding pebble radius is estimated by
\begin{equation}
r_\mathrm{bounce} = \left(\frac{3m_\mathrm{bounce}}{4\pi\rho_{0,\mathrm{ice+sil}}\phi_\mathrm{p}}\right)^{1/3}.
\end{equation}
{\revisetwo{The pebble radius as a function of $\phi_\mathrm{p}$ is shown in the bottom panel of Figure \ref{fig:phi_p_depend}, which suggests that $r_\mathrm{bounce}\lesssim 130\mathrm{\ \mu m}$, although it has a dependence on the tensile strength of 67P and $\phi_\mathrm{p}$.}}

{\reviseone{It is worth noting that the predicted pebble sizes are comparable to the maximum sizes of pebbles limited by the bouncing barrier in protoplanetary disks \citep{Dominik2024,Oshiro2025}.
The predicted pebble sizes are also consistent with millimeter polarimetric observations of protoplanetary disks, which typically estimate maximum grain sizes around $100\mathrm{\ \mu m}$ \citep[e.g.,][]{Kataoka2016HLTau,Kataoka2016HD,Kataoka2017,Bacciotti2018,Liu2019,Miotello2023,Drazkowska2023,Ueda2024}.}}
However, the estimate of $r_\mathrm{bounce}$ is highly sensitive to the volume filling factor $\phi_\mathrm{p}$, and thus uncertainties in $\phi_\mathrm{p}$ can cause large variations in $r_\mathrm{bounce}$.

\section{Conclusions} \label{sec:conclusion}

We developed a new model of the contact surfaces between colliding pebbles, based on the formulation of the compressive strength of dust aggregates \citep{Tatsuuma2023}, and validated our model through numerical simulations.
Using our model, we examined the formation process of comet 67P.

{\reviseone{The relationship between the collision velocity $v_0$ and the contact surface radius normalized by pebble radius, $a/r_\mathrm{p}$, is given by Equations (\ref{eq:phi_e}) and (\ref{eq:v_0}).
In this model, we assume that the volume filling factor increases uniformly from $\phi_\mathrm{s}$ to $\phi_\mathrm{e}$ within the compressed region, whose volume decreases from $Ca^3+V_\mathrm{r}$ to $Ca^3$ with $C=1.5$.
This model also assumes that the initial kinetic energy of the pebbles is entirely used for inelastic compression.
We confirmed that our model reproduces numerical simulation results, in which both the contact surface and pebble radii are measured using two- and three-dimensional characteristic radii given by Equations (\ref{eq:chara2d}) and (\ref{eq:chara3d}), respectively.
Due to internal fragmentation, the model is not applicable to silicate pebbles with an initial volume filling factor above 0.3.
In addition, the number of contact points on the contact surface, which also plays a key role in determining mechanical strength and thermal conductivity, is given by Equations (\ref{eq:phi_e}) and (\ref{eq:N_con}) with $C_\mathrm{con}=0.72$.}}

{\reviseone{Using the tensile strength of 67P to constrain the contact surface radius between pebbles, we inferred the collision velocities that must have occurred during its formation.
The contact surface radius normalized by pebble radius $a/r_\mathrm{p}$ needs to be 0.002--0.09 (Equation (\ref{eq:67Pa_rp})), which suggests that 67P was formed via pebble collisions at low collision velocities {\revisethree{below $\sim10\mathrm{\ cm\ s^{-1}}$ when a microscopic filling factor of pebbles is lower than 0.6 (Figure \ref{fig:phi_p_depend}).}}}}

{\reviseone{We proposed a scenario in which 67P formed from a gravitationally collapsing cloud of inelastically bouncing pebbles.
Based on this scenario and the recent simulation results by \citet{Oshiro2025}, we predicted that the ice pebbles that formed 67P have radii of 130 $\mathrm{\mu m}$ or smaller, assuming a microscopic filling factor of 0.5 or higher.
The predicted pebble size is consistent with the maximum size set by the bouncing barrier in protoplanetary disks and with observationally inferred pebble sizes in some disks.}}

{\reviseone{In this work, we suggested the process of pebble coagulation by constraining the physical properties of pebbles using the bulk properties of comet 67P.
We found that inelastic bouncing within gravitationally bound pebble clouds plays a crucial role.
It should be noted that the estimate of pebble size from the bouncing scenario is highly sensitive to the microscopic filling factor of pebbles.
We still need to confirm the scenario of inelastic bouncing through direct simulations of gravitationally collapsing pebble clouds, which are thought to be formed via streaming instability.}}

\begin{acknowledgments}

We thank Sota Arakawa and Haruto Oshiro for fruitful discussions.
This work was supported by JSPS KAKENHI grant Nos. JP22J00260, JP22KJ1292, JP23K25923, and JP25K23408.
This work has made use of NASA’s Astrophysics Data System.
This work has made use of adstex (https://github.com/yymao/adstex).

\end{acknowledgments}

\appendix

\section{Comparison with Alternative Model of Contact Surface Radius}
\label{apsec:anothermodel}

In this section, we introduce an alternative model, referred to as model 2, for the contact surface radius based on the assumption that the compressed region decreases from $C_2a^2\delta+V_\mathrm{r}$ to $C_2a^2\delta$.
We also determine the minimum value of $\sigma^2$, identify the best-fitting $C_2$, and evaluate them.

In model 2, we assume that
\begin{equation}
(C_2a^2\delta+V_\mathrm{r})\phi_\mathrm{s} = C_2a^2\delta\phi_\mathrm{e}
\end{equation}
instead of Equation (\ref{eq:volume-filling}).
The thickness of the compressed region, $\delta$, is given by
\begin{eqnarray}
\delta &=& r_\mathrm{p}-\sqrt{r_\mathrm{p}^2-a^2},\nonumber\\
\frac{\delta}{r_\mathrm{p}} &=& 1-\sqrt{1-\left(\frac{a}{r_\mathrm{p}}\right)^2}.
\end{eqnarray}
The volume filling factor after compression is given by
\begin{eqnarray}
\phi_\mathrm{e} &=& \left\{1+\frac{4\pi}{3C_2}\left(\frac{a}{r_\mathrm{p}}\right)^{-2}\left(\frac{\delta}{r_\mathrm{p}}\right)^{-1}\right.\nonumber\\
&&-\left.\frac{2\pi}{3C_2}\left(\frac{\delta}{r_\mathrm{p}}\right)^{-1}\left[2\left(\frac{a}{r_\mathrm{p}}\right)^{-2}+1\right]\right.\nonumber\\
&&\times\left.\sqrt{1-\left(\frac{a}{r_\mathrm{p}}\right)^2}\right\}\phi_\mathrm{s}.
\label{eq:phi_e_an}
\end{eqnarray}
Finally, we obtain
\begin{eqnarray}
v_0^2 &=& 
{\frac{2C_2E_\mathrm{roll}}{m_0}}\frac{\phi_\mathrm{e}}{\phi_\mathrm{s}}\left(\frac{a}{r_\mathrm{p}}\right)^2\frac{\delta}{r_\mathrm{p}}\nonumber\\
&&\times\left[\left(\frac{1}{\phi_\mathrm{e}}-\frac{1}{\phi_\mathrm{max}}\right)^{-2}-\left(\frac{1}{\phi_\mathrm{s}}-\frac{1}{\phi_\mathrm{max}}\right)^{-2}\right].
\label{eq:v_0_an}
\end{eqnarray}

\begin{figure*}[t!]
\plotone{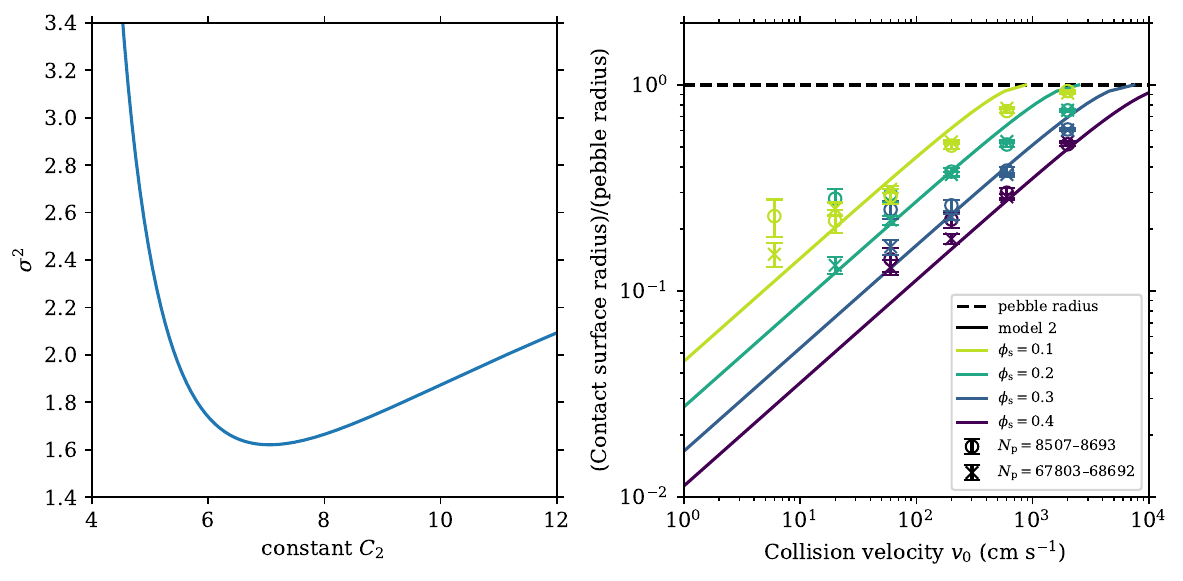} 
\caption{Left: values of $\sigma^2$ against the constant $C$, calculated using Equation (\ref{eq:chi2}).
The minimum value is $\sigma^2=1.6$ when $C_2=7.1$.
Right: relationship between the collision velocity $v_0$ and $a/r_\mathrm{p}$ for ice pebbles and model 2.
The solid lines represent model 2 (Equations (\ref{eq:phi_e_an}) and (\ref{eq:v_0_an})).
The other lines, markers, and colors are the same as in Figure \ref{fig:velocity_contact_area}.
\label{fig:model_delta}}
\end{figure*}

As in Appendix \ref{apsec:fitting}, we calculate $\sigma^2$ over a certain range of $C_2$ to determine the best-fitting $C_2$ using Equation (\ref{eq:chi2}).
Figure \ref{fig:model_delta} shows the values of $\sigma^2$ and indicates that the minimum value is $\sigma^2=1.6$ when $C_2=7.1$.
For the best-fitting $C_2$ and ice pebbles, we plot the relationship between the collision velocity $v_0$ and $a/r_\mathrm{p}$ from our simulation results, as well as that from model 2 (Equations (\ref{eq:phi_e_an}) and (\ref{eq:v_0_an})), in Figure \ref{fig:model_delta}.
The assumption of $Ca^3$ is more accurate than $C_2a^2\delta$ because the minimum value for the former model ($\sigma^2=0.62$) is smaller than that for the latter model ($\sigma^2=1.6$).

\section{Monte Carlo Error Estimation for {\lowercase{$a_\mathrm{c,2d}$}}}
\label{apsec:errorMC}

In this section, we estimate the error in the contact surface radius using the Monte Carlo method.
We conduct 100 Monte Carlo simulations for each given number of particles $N_\mathrm{par}=3$, 10, 30, 100, 300, 1000, 3000, and 10000.
In each simulation, $N_\mathrm{par}$ particles are randomly distributed within a circle of radius $R_\mathrm{out}=0.5$, centered at (0.5, 0.5), and the two-dimensional characteristic radius $a_\mathrm{c,2d}$ is calculated with Equation (\ref{eq:chara2d}).
We obtain the mean and standard deviation of $a_\mathrm{c,2d}$ from 100 simulations for each value of $N_\mathrm{par}$.
We summarize the Monte Carlo simulation results in Table \ref{tab:error}.

The error in the contact surface radius is proportional to $1/\sqrt{N_\mathrm{par}}$.
From Table \ref{tab:error}, we obtain the error of $a_\mathrm{c,2d}/R_\mathrm{out}$ given as
\begin{equation}
\sigma_\mathrm{c,2d} = \frac{0.29}{\sqrt{N_\mathrm{par}}}.
\label{eq:error}
\end{equation}

\begin{deluxetable}{ccc}
\tablecaption{Results of Monte Carlo Error Estimates for the Two-dimensional Characteristic Radius $a_\mathrm{c,2d}$.
\label{tab:error}}
\tablehead{
\colhead{$N_\mathrm{par}$} & \colhead{Mean of $a_\mathrm{c,2d}/R_\mathrm{out}$} & \colhead{Standard deviation of $a_\mathrm{c,2d}/R_\mathrm{out}$}
}
\startdata 
3 & 0.97 & 0.17 \\
10 & 0.993 & 0.092 \\
30 & 1.004 & 0.052 \\
100 & 0.998 & 0.029 \\
300 & 1.001 & 0.015 \\
1000 & 1.000 & 0.010 \\
3000 & 1.0004 & 0.0054 \\
10000 & 1.0003 & 0.0028 \\
\enddata
\tablecomments{$N_\mathrm{par}$ represents the number of particles inside the circle, and $R_\mathrm{out}=0.5$ is the radius of the circle.}
\end{deluxetable}

\section{Fitting Methods and Results}
\label{apsec:fitting}

In this section, we describe the fitting methods for the constants $C$ and $C_\mathrm{con}$ in Equations (\ref{eq:v_0}) and (\ref{eq:N_con}), respectively, along with the results.

\begin{figure}[t!]
\plotone{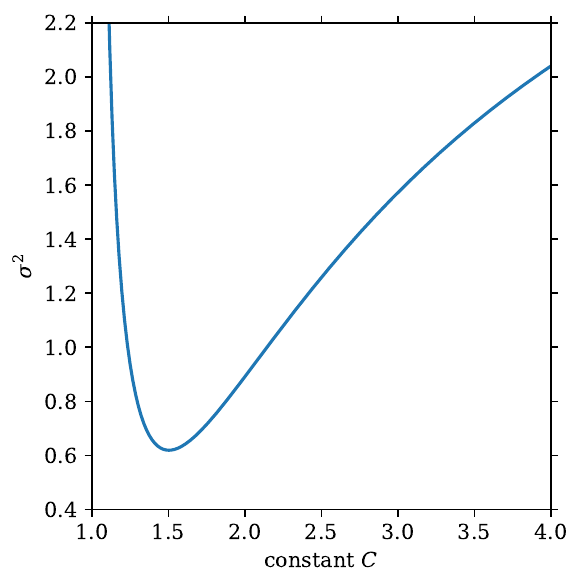} 
\caption{Values of $\sigma^2$ against the constant $C$, calculated using Equation (\ref{eq:chi2}).
The minimum value is $\sigma^2=0.62$ when $C=1.5$.
\label{fig:constant_chi2}}
\end{figure}

We calculate $\sigma^2$ over a certain range of $C$ to find the best-fitting $C$.
For each $a_\mathrm{data}/r_\mathrm{p,data}$ obtained from simulations, we use both the corresponding simulation results $v_\mathrm{0,data}$ and $v_0(a_\mathrm{data}/r_\mathrm{p,data})$, which is obtained by substituting $a_\mathrm{data}/r_\mathrm{p,data}$ into Equations (\ref{eq:phi_e}) and (\ref{eq:v_0}).
We calculate $\sigma^2$ given as
\begin{equation}
\sigma^2 = \sum\left[\log_{10}v_\mathrm{0,data}-\log_{10}v_0\left(\frac{a_\mathrm{data}}{r_\mathrm{p,data}}\right)\right]^2.
\label{eq:chi2}
\end{equation}
Figure \ref{fig:constant_chi2} shows the values of $\sigma^2$ and shows that the minimum value is $\sigma^2=0.62$ when $C=1.5$.

\begin{figure}[b!]
\plotone{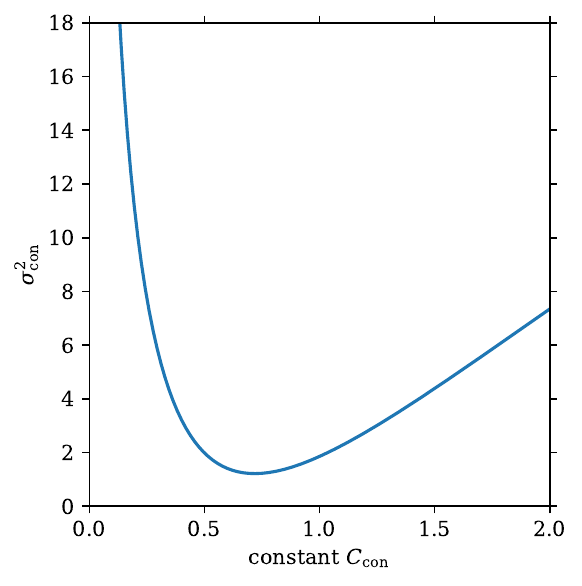} 
\caption{Values of $\sigma_\mathrm{con}^2$ against the constant $C_\mathrm{con}$, calculated using Equation (\ref{eq:sigma_con2}).
The minimum value is $\sigma_\mathrm{con}^2=1.2$ when $C_\mathrm{con}=0.72$.}
\label{fig:constant_con_chi2}
\end{figure}

Similarly, we calculate $\sigma_\mathrm{con}^2$ over a range of $C_\mathrm{con}$ to determine the best-fitting $C_\mathrm{con}$.
For each $a_\mathrm{data}/r_\mathrm{p,data}$ obtained from simulations, we use both the corresponding simulation results $N_\mathrm{con,data}$ and $\phi_\mathrm{e}(a_\mathrm{data}/r_\mathrm{p,data})$ obtained by substituting $a_\mathrm{data}/r_\mathrm{p,data}$ into Equation (\ref{eq:phi_e}).
We calculate $\sigma_\mathrm{con}^2$ as
\begin{eqnarray}
\sigma_\mathrm{con}^2 = &\sum&\left[\log_{10}\frac{N_\mathrm{con,data}}{(r_\mathrm{p,data}/r_0)^2}\right.\nonumber\\
&&-\left.\log_{10}C_\mathrm{con}\phi_\mathrm{e}\left(\frac{a_\mathrm{data}}{r_\mathrm{p,data}}\right)\left(\frac{a_\mathrm{data}}{r_\mathrm{p,data}}\right)^2\right]^2.
\label{eq:sigma_con2}
\end{eqnarray}
Figure \ref{fig:constant_con_chi2} shows the values of $\sigma_\mathrm{con}^2$ and indicates that the minimum value is $\sigma_\mathrm{con}^2=1.2$ when $C_\mathrm{con}=0.72$.

\section{Power Law at Low Collision Velocities}
\label{apsec:modelinterp}

In this section, we derive the power law for the collision velocity $v_0$ at low velocities by assuming $a\ll r_\mathrm{p}$.
By neglecting the higher-order terms in $a/r_\mathrm{p}$, Equation (\ref{eq:phi_e}) simplifies to
\begin{eqnarray}
\phi_\mathrm{e} &\simeq& \left\{1+\frac{4\pi}{3C}\left(\frac{a}{r_\mathrm{p}}\right)^{-3}\right. \nonumber\\
&&-\left.\frac{2\pi}{3C}\left[2\left(\frac{a}{r_\mathrm{p}}\right)^{-2}+1\right]\left[\left(\frac{a}{r_\mathrm{p}}\right)^{-1}-\frac{1}{2}\left(\frac{a}{r_\mathrm{p}}\right)\right]\right\}\phi_\mathrm{s} \nonumber \\
&=& \left[1+\frac{\pi}{3C}\left(\frac{a}{r_\mathrm{p}}\right)\right]\phi_\mathrm{s}.
\label{eq:phi_e_simple}
\end{eqnarray}
Using
\begin{eqnarray}
\left(\frac{1}{\phi_\mathrm{e}}-\frac{1}{\phi_\mathrm{max}}\right)^{-2} 
&\simeq& \left(\frac{1}{\phi_\mathrm{s}}-\frac{1}{\phi_\mathrm{max}}\right)^{-2}\nonumber\\
&&+ \frac{2\pi}{3C\phi_\mathrm{s}}\left(\frac{1}{\phi_\mathrm{s}}-\frac{1}{\phi_\mathrm{max}}\right)^{-3}\frac{a}{r_\mathrm{p}},
\end{eqnarray}
Equation (\ref{eq:v_0}) becomes
\begin{eqnarray}
v_0
&\simeq& 2\sqrt{\frac{\pi E_\mathrm{roll}}{3m_0}\phi_\mathrm{e}\phi_\mathrm{s}\left(1-\frac{\phi_\mathrm{s}}{\phi_\mathrm{max}}\right)^{-3}}\left(\frac{a}{r_\mathrm{p}}\right)^2 \nonumber \\
&\simeq& 2\sqrt{\frac{\pi E_\mathrm{roll}}{3m_0}\left(1-\frac{\phi_\mathrm{s}}{\phi_\mathrm{max}}\right)^{-3}}\left(\frac{a}{r_\mathrm{p}}\right)^2\phi_\mathrm{s},\\
\frac{a}{r_\mathrm{p}} &=& \left(\frac{3m_0}{4\pi E_\mathrm{roll}}\right)^{1/4}\left(1-\frac{\phi_\mathrm{s}}{\phi_\mathrm{max}}\right)^{3/4}\left(\frac{v_0}{\phi_\mathrm{s}}\right)^{1/2}.
\label{eq:v_0_powerlaw}
\end{eqnarray}

\newpage 

\begin{figure}[t!]
\plotone{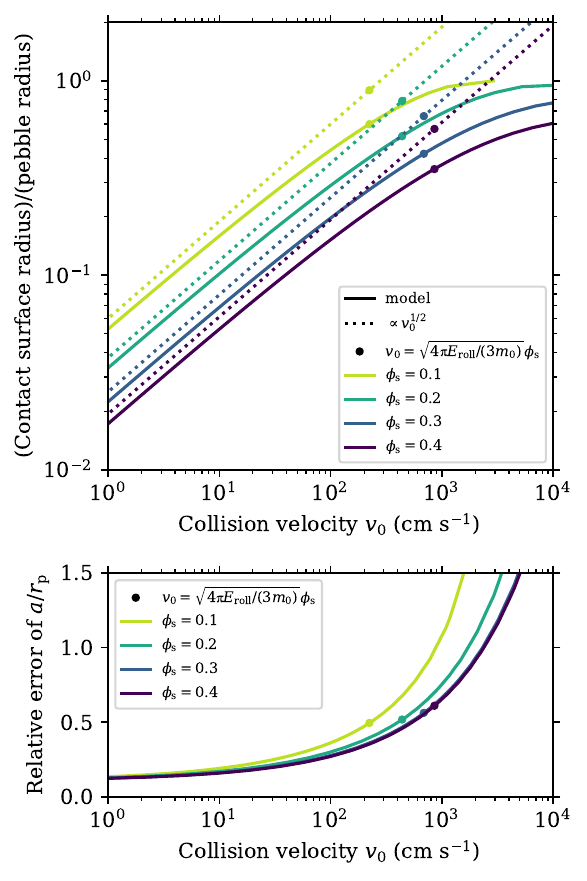} 
\caption{Top: relationship between the collision velocity $v_0$ and $a/r_\mathrm{p}$ for ice pebbles.
The solid lines and colors are the same as in Figure \ref{fig:velocity_contact_area}.
The dotted lines show the power law of $(a/r_\mathrm{p})\propto v_0^{1/2}$ under the assumption that $a\ll r_\mathrm{p}$.
The circles represent $v_0=\sqrt{4\pi E_\mathrm{roll}/(3m_0)}\phi_\mathrm{s}$.
Bottom: relative error to our model (Equations (\ref{eq:phi_e}) and (\ref{eq:v_0})).
\label{fig:model_compare}}
\end{figure}

Figure \ref{fig:model_compare} shows that the relationship at low collision velocities is described by the power law of $(a/r_\mathrm{p})\propto v_0^{1/2}$.
The relative error of $a/r_\mathrm{p}$ with respect to our model (Equations (\ref{eq:phi_e}) and (\ref{eq:v_0})) does not exceed $\simeq 0.5$ if
\begin{equation}
v_0 \leq \sqrt{\frac{4\pi E_\mathrm{roll}}{3m_0}}\phi_\mathrm{s}.
\end{equation}

\software{VisIt \citep{VisIt}}

\bibliography{paper}{}
\bibliographystyle{aasjournal}

\end{document}